\documentclass[showpacs,amsmath,amssymb,aps,showkeys,floatfix,prd,a4paper]{revtex4}

\usepackage[dvips]{graphicx}
\usepackage{dcolumn}
\usepackage{bm}
\usepackage{epsfig}
\usepackage{amsfonts}
\usepackage{amssymb,amscd}

\def\lsim{\raise0.3ex\hbox{$<$\kern-0.75em\raise-1.1ex\hbox{$\sim$}}}

\def\gsim{\raise0.3ex\hbox{$>$\kern-0.75em\raise-1.1ex\hbox{$\sim$}}}

\def\pom{{I\!\!P}}

\newcommand{\be}{\begin{equation}}

\newcommand{\ee}{\end{equation}}

\def\beq{\begin{equation}}

\def\eeq{\end{equation}}

\def\beqa{\begin{eqnarray}}

\def\eeqa{\end{eqnarray}}

\newcommand{\rd}{\mbox{\boldmath $\Delta$}}

\newcommand{\ba}{\begin{eqnarray}}

\newcommand{\rr}{\mbox{\boldmath $r$}}

\newcommand{\rb}{\mbox{\boldmath $b$}}

\def\gappeq{\mathrel{\rlap {\raise.5ex\hbox{$>$}}

{\lower.5ex\hbox{$\sim$}}}}

\def\lappeq{\mathrel{\rlap{\raise.5ex\hbox{$<$}}

{\lower.5ex\hbox{$\sim$}}}}

\def\Toprel#1\over#2{\mathrel{\mathop{#2}\limits^{#1}}}

\def\pom{{I\!\!P}}

\begin{document}


\title{Associated vector meson and bound-free electron-positron pair photoproduction \\ in ultraperipheral $PbPb$ collisions\footnote{This paper is dedicated to the memory of Prof. Gerard Baur, one of the founders of the UPC physics.}}

\author{Celsina N. {\sc Azevedo}}
\email{acelsina@gmail.com}
\affiliation{ Institute of Physics and Mathematics, Federal University of Pelotas, \\
  Postal Code 354,  96010-900, Pelotas, RS, Brazil}

\author{Victor P. {\sc Gon\c{c}alves}}
\email{barros@ufpel.edu.br}
\affiliation{ Institute of Physics and Mathematics, Federal University of Pelotas, \\
  Postal Code 354,  96010-900, Pelotas, RS, Brazil}
\affiliation{Institute of Modern Physics, Chinese Academy of Sciences,
  Lanzhou 730000, China}

\author{Bruno D. {\sc Moreira}}
\email{bduartesm@gmail.com}
\affiliation{Departamento de F\'isica, Universidade do Estado de Santa Catarina, 89219-710 Joinville, SC, Brazil.}

\begin{abstract}
In this letter we analyze the associated production of a vector meson with the bound - free $e^+e^-$ process in ultraperipheral $PbPb$ collisions through the double scattering mechanism for the energy of the Large Hadron Collider (LHC). Such process is characterized by the presence of a meson and a positron in the final state and by a forward hydrogen - like ion with a distinct electric charge. We present our predictions for the total cross sections and rapidity distributions considering the rapidity ranges covered by the ALICE and LHCb detectors, which indicate that a future experimental analysis of the $\phi + e^+$ and  $J/\Psi + e^+$ final states is feasible. 
\end{abstract}



\maketitle

\vspace{1cm}


Relativistic heavy - ion collisions at the Large Hadron Collider (LHC) induce strong electromagnetic fields, which may lead to the particle production by photon - photon and photon - hadron interactions. In particular, for ultraperipheral collisions, characterized by an impact parameter larger than the sum of the radii of the incident ions, the strong interactions are suppressed, which allow us to consider the  LHC as a photon collider \cite{upc1,upc2}.  Over the last decades, a large number of  experimental and theoretical studies have been performed, mainly focused on  the dilepton production by photon -- photon ($\gamma \gamma$) interactions and vector mesons by photon -- Pomeron ($\gamma \pom$) interactions, mainly motivated by the large cross sections associated with these processes and by the perspective of probing the basic QED reactions (as e.g. the Light - by - Light scattering) and improving our understanding about the QCD dynamics (For a recent review see, e.g. Ref. \cite{upc2}). Another electromagnetic process of interest for heavy - ion colliders is the bound - free electron - positron pair production (BFPP), in which the electron produced in the $\gamma \gamma \rightarrow e^+e^-$ subprocess is captured by one of the incident nuclei, creating a hydrogen-like ion \cite{Baur:2007zz}. The capture implies a change in the magnetic rigidity that leads to the loss of the ion from the beam, which will be lost in a well - defined spot in the collider ring. As a consequence, the BFPP process contributes to the intensity and luminosity decay of heavy - ion colliders and such an aspect has motivated a series of studies \cite{Baltz:1991zz,Baltz:1993zz,Belkacem:1993wq,Aste:1994fjr,
Meier:2000ga,Lee:2000ap,Aste:2007vs,Bruce:2007zza,Bruce:2009bg,Sengul:2011zz,Artemyev:2012ja,Artemyev:2014eaa,Bauer:2020hsr,Schaumann:2020lsl,Bruce:2021hii}.

\begin{figure}[t]
\includegraphics[scale=0.4]{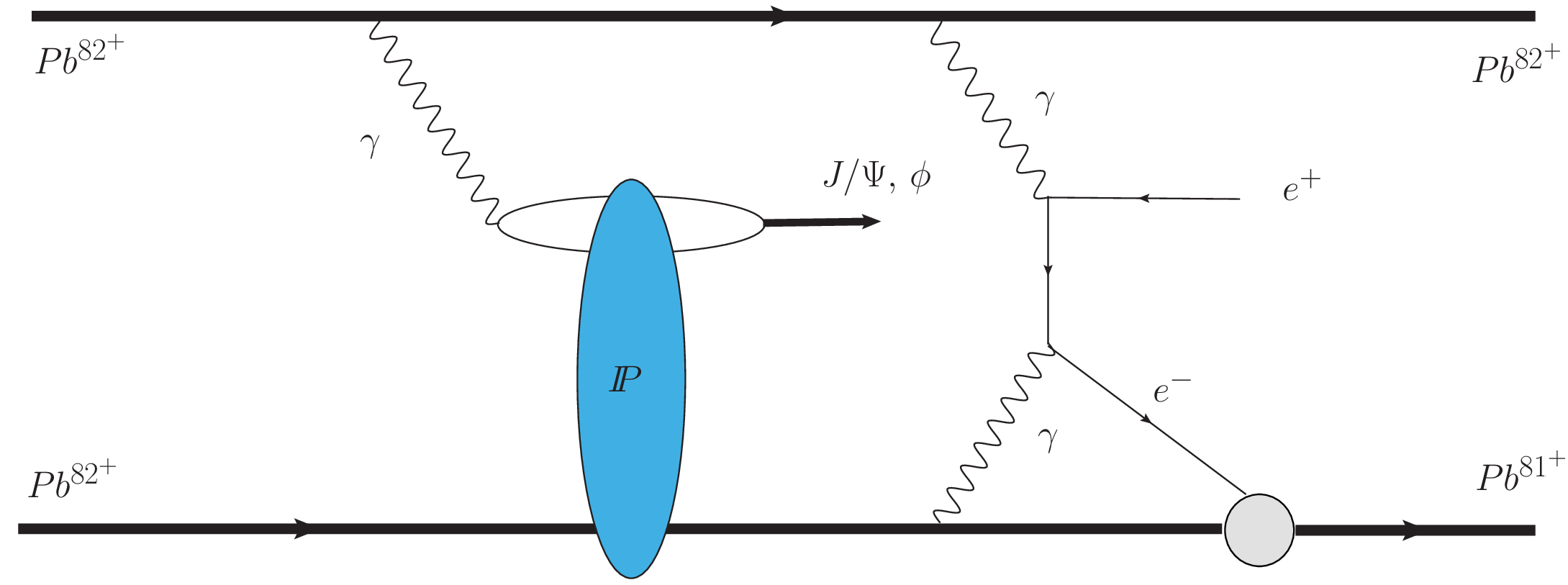}
\caption{Associated exclusive vector meson and bound - free dielectron photoproduction in ultraperipheral $Pb Pb$ collisions.}
\label{Fig:diagram}
\end{figure}

In recent years, several studies have demonstrated that the high photon luminosity present in ultraperipheral heavy ion collisions implies a non - negligible contribution of the  double particle production process in photon - induced interactions, which can be considered as an alternative to improve our understanding of the QCD dynamics as well as can be used for testing the treatment of the double scattering mechanism in UPHICs \cite{klein,Klusek-Gawenda:2013dka,Artemyev:2012ja,Artemyev:2014eaa,DSM,Azevedo:2022ozu,Klusek-Gawenda:2016suk,vanHameren:2017krz,Karadag:2019gvc,Azevedo:2023vsz}.
In particular, the detailed discussion presented in Ref. \cite{Artemyev:2014eaa} about the double lepton pair production with electron capture in ultrarelativistic heavy-ion collisions indicated that the observation of this process may become feasible at the LHC. Its experimental separation  was discussed in Ref. \cite{Schicker:2015mmj} and  the measurement of the forward scattered $Pb$ ion of different charge was suggested as a way to tag the associated dilepton production at midrapidity in the central detector. Moreover, Ref. \cite{Schicker:2015mmj} has also pointed out the interesting possibility of measuring the associated production of a  meson pair and a bound - free $e^+e^-$ pair, but  an estimate for the magnitude of the corresponding cross section was not presented. Our goal in this letter is to analyze in more detail such a suggestion and provide, for the first time, predictions for the rapidity distributions and total cross sections. However, in contrast with Ref. \cite{Schicker:2015mmj}, we will consider the exclusive $J/\Psi$ and $\phi$ photoproduction by $\gamma \pom$ interactions in association with the BFPP process, as represented in Fig. \ref{Fig:diagram}. Such changing is motivated by the following aspects: (a) the magnitude of the  cross section for the exclusive vector meson photoproduction at high energies is larger than that for the $\gamma \gamma \rightarrow M \bar{M}$ process, where $M$ is a meson; and (b) the experimental separation of a single vector meson in the final state is easier than the measurement of the meson pair. In this exploratory study, we will assume that the $Pb^{81^+}$ ion can be tagged by a forward detector and the vector meson is measured in the central detector. Predictions for $PbPb$ collisions at the LHC energy will be presented considering the rapidity ranges covered by the central and forward detectors. It is important to emphasize that in this letter we will not impose any cut on the transverse momentum and rapidity of the positron.  However, a very interesting possibility, that deserves a more detailed study, is that  the positron  is also observed in the central detector. In principle, in such a case, the cross sections will be smaller than the values presented in this paper, but the topology of the final state would be a clear signature of the double scattering mechanism. Such an alternative will be analyzed in a forthcoming study.

Initially, let's present a brief review of the formalism needed to describe the associated vector meson and bound - free dielectron photoproduction via double scattering mechanism (DSM) in UPHICs (For details see, e.g., Refs. \cite{klein,DSM}). As in previous studies \cite{klein,Klusek-Gawenda:2013dka,Artemyev:2012ja,Artemyev:2014eaa,DSM,Azevedo:2022ozu,Klusek-Gawenda:2016suk,vanHameren:2017krz,Karadag:2019gvc,Azevedo:2023vsz}, we will assume  that possible correlations  can be neglected. Such approximation allow us to express  the pair production probability  in terms of the product of single production probabilities for the exclusive vector meson photoproduction and for the BFPP process. As we will not impose any restriction on the rapidity of the positron, we will integrated over all possible values for this quantity. As a consequence, the differential cross section for the  production of a vector meson $V$ $(= J/\Psi,\,\phi)$  at rapidity $Y_{V}$ in association with the BFPP process will be given by \cite{klein,Klusek-Gawenda:2013dka,DSM}
\begin{eqnarray}
\frac{d\sigma \left[Pb^{82^+} Pb^{82^+}  \rightarrow   Pb^{82^+} \,\, {{V}} \, {e^+} \,\, Pb^{81^+} \right]}{dY_{{{V}}}} & = &  \int_{b_{min}}
\frac{d\sigma \,\left[Pb^{82^+} Pb^{82^+} \rightarrow   Pb^{82^+} \otimes {{V}} \otimes Pb^{82^+} \right]}{d^2{\mathbf b} dY_{{{V}}}} \nonumber\\
&  \times  &
\frac{d\sigma \,\left[Pb^{82^+} Pb^{82^+} \rightarrow   Pb^{82^+} \otimes  {e^+} \otimes Pb^{81^+} \right]}{d^2{\mathbf b}} \,d^2{\mathbf b} \,\,,
\label{Eq:double}
\end{eqnarray} 
where ${\mathbf b}$ is the impact parameter of the collision and $\otimes$ represents the presence of a rapidity gap in the final state. In our analysis we will assume  $b_{min} = 2 R_{Pb}$,  which is equivalent to treat 
the nuclei as hard spheres and that  excludes the overlap between the colliding hadrons. Similar predictions are  obtained assuming  $b_{min} = 0$ and including the survival factor $P_{NH}(b)$ that describes the probability of no additional hadronic interaction between the nuclei, which is usually estimated using the Glauber formalism \cite{klein,Baltz:2009jk}.  The differential cross section for the exclusive vector meson photoproduction 
can be expressed by
\begin{eqnarray}
\frac{d\sigma \,\left[Pb^{82^+} Pb^{82^+} \rightarrow   Pb^{82^+} \otimes  {V} \otimes Pb^{82^+}\right]}{d^2{\mathbf b} dY_{V}} = \omega N_{Pb^{82^+}}(\omega,{\mathbf b})\,\sigma_{\gamma Pb^{82^+} \rightarrow {V} \otimes Pb^{82^+}}\left(\omega \right)\,\,,
\label{dsigdy}
\end{eqnarray}
where the rapidity $Y_{V}$ of the vector meson in the final state is determined by the photon energy $\omega$ in the collider frame and by mass $M_{V}$ of the vector meson [$Y_{V}\propto \ln \, ( \omega/M_{V})$]. Moreover, $N_{Pb^{82^+}}(\omega,{\mathbf b})$ is  the equivalent photon spectrum associated with the ${Pb^{82^+}}$ ion, which can be expressed in terms of the nuclear form factor. In our analysis, we will estimate the photon flux assuming the realistic form factor, which corresponds to the Wood - Saxon distribution and is the Fourier transform of the charge density of the nucleus, constrained by the experimental data. Moreover, the total cross section for the exclusive vector meson photoproduction, $\sigma_{\gamma {Pb^{82^+}} \rightarrow {V} \otimes {Pb^{82^+}}}$,  
will be estimated using the color dipole formalism, assuming the Gaus-LC model for the overlap function and the Glauber - Gribov model for the non-forward scattering  amplitude (See Refs. \cite{DSM,Azevedo:2022ozu} for details). As demonstrated e.g. in  Ref. \cite{run2}, such a formalism describes the current data for the photoproduction of vector mesons in UPHICs. On the other hand, the differential cross section for the bound - free dielectron production will be estimated using the formalism described in detail in Ref. \cite{Sengul:2011zz}. The basic idea is that after the creation of the $e^+e^-$ pair, the electron is captured by one of the incident ions and then the positron becomes free. The BFPP cross section will be calculated in lowest - order QED using the semiclassical approximation and assuming the Sommerfeld - Maue and Darwin wave - functions for the positron and for the captured electron, respectively. As demonstrated in Ref. \cite{Sengul:2011zz}, the resulting predictions are similar to those derived in Refs. \cite{Baltz:1991zz,Baltz:1993zz,Aste:1994fjr,
Meier:2000ga,Aste:2007vs} using distinct approaches.

\begin{figure}[t]
\includegraphics[scale=0.48]{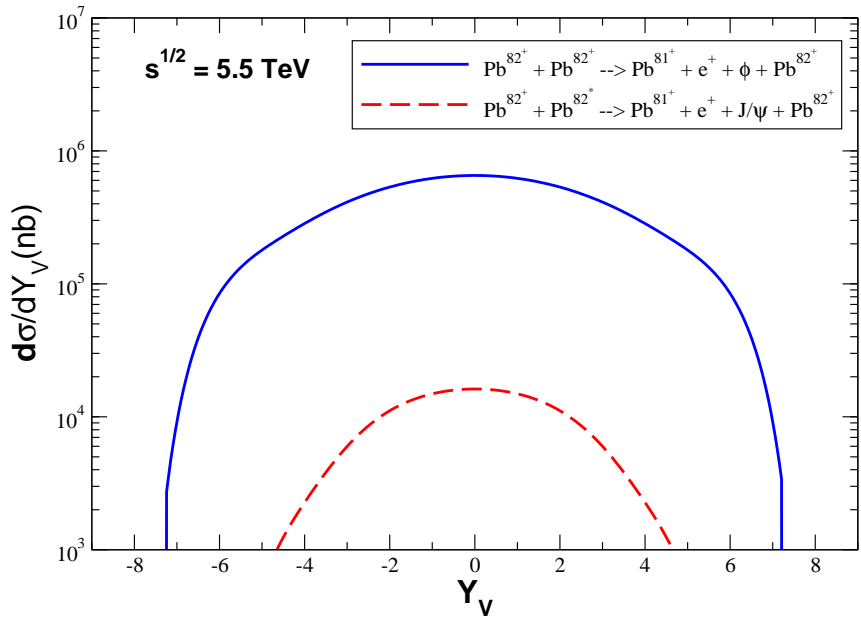} 
\caption{Rapidity distribution of the vector mesons derived assuming that these are produced in association with the bound - free $e^+e^-$ process via DSM in ultraperipheral $PbPb$ collisions at the LHC energy. Results are derived integrating over the positron rapidity.}
\label{Fig:rapidity}
\end{figure}

\begin{table}[t]
\begin{center}
\begin{tabular}{||c|c|c|c||}\hline \hline
                                                & Full rapidity range   & $-2.5 \le Y_V \le 2.5$        & $2.0 \le Y_V \le 4.5$   \\ \hline \hline                                                
$Pb^{82^+} + Pb^{82^+} \rightarrow Pb^{81^+} + e^+ + J/\Psi + Pb^{82^+}$     & $5.74 \times 10^{3}$  & $4.53 \times 10^{3}$  & $8.92 \times 10^{2}$ \\ \hline
$Pb^{82^+} + Pb^{82^+} \rightarrow Pb^{81^+} + e^+ + \phi + Pb^{82^+}$  & $2.85 \times 10^{5}$  & $1.72 \times 10^{5}$  & $5.53 \times 10^{4}$ \\ \hline \hline
\end{tabular} 
\caption{Total cross sections (in nb) for the associated production of $J/\Psi$ and 
$\phi$ with bound-free $e^{+}e^{-}$ pairs in ultraperipheral $PbPb$ collisions 
(with $\sqrt{s} = 5.5$ TeV) considering distinct rapidity ranges for the vector meson.}
\label{Tab:secao-de-choque}
\end{center}
\end{table}

In Fig. \ref{Fig:rapidity} we present our  predictions for the rapidity distribution of the vector mesons derived assuming that these are produced in association  with the BFPP process via DSM in ultraperipheral $PbPb$ collisions at the LHC energy ($\sqrt{s} = 5.5$ TeV). 
One has that predictions involving a $J/\Psi$ meson have a smaller normalization and are narrower in rapidity than those with a $\phi$ meson. In general, the normalization for midrapidities is a factor $\gtrsim 50$ smaller. In comparison to the predictions presented e.g. in Ref. \cite{run2} for the exclusive vector meson photoproduction via single scattering mechanism, i.e. without the presence of the BFPP process, one has that the DSM predictions are a factor $\gtrsim 200$  smaller.

The total cross sections for the associated vector meson and bound - free dielectron photoproduction via DSM in ultraperipheral $PbPb$ collisions are presented in Table \ref{Tab:secao-de-choque} considering the LHC  energy as well as different ranges for the rapidity of the vector meson. We present predictions for the full rapidity range of LHC, as well as assuming that the vector meson is detected by a central ($-2.5  \le Y_{V} \le 2.5$) or a forward ($2.0 \le Y_V \le 4.5$) detector, as e.g. the ALICE and LHCb detectors, respectively. One has that the prediction for the  $\phi$ + BFPP process is almost two orders of magnitude larger than those obtained for the $J/\Psi$  + BFPP case. In comparison with the results presented in Ref. \cite{Azevedo:2023vsz}, where the cross section for the associated production of a vector meson with a free dielectron pair  was estimated, our predictions for the $V$ + BFPP case are smaller by a factor $\gtrsim 20$. Assuming that the  integrated luminosity expected for future runs of heavy - ion collisions at the LHC  will be  ${\cal{L}} \approx 7$ nb$^{-1}$, the  number of $\phi$ + BFPP events by year  will be $\ge 10^5 \, (10^4)$ considering the central (forward) rapidity range. On the other hand, for the  
$J/\Psi$ + BFPP case, our predictions are smaller by approximately two orders of magnitude. Such results indicate that  a future experimental of the exclusive vector meson photoproduction with the BFPP process is, in principle, feasible.


As a summary,  over the last decades, several studies have demonstrated that the cross sections for photon - induced interactions in ultraperipheral heavy - ion collisions are huge and can be used to improve our understanding of the QCD dynamics at high energies,  to probing the basic QED processes and searching for New Physics. In particular, recent studies indicated that the double particle production via the double scattering mechanism in ultraperipheral heavy ion collisions is non - negligible and can be considered an alternative to study photon - induced processes.
Such results have motivated the analysis performed in this letter, where we have estimated, for the first time, the associated production of a vector meson with a bound - free dielectron process in ultraperipheral $PbPb$ collisions for the LHC energy and presented predictions for the total cross sections and rapidity distributions considering the phase space covered by the ALICE and LHCb Collaborations. We predict large values for the total cross sections and for the number of events in future runs of the LHC, which indicate that the study of these final states are, in principle, feasible. The results presented in this exploratory study strongly motivate a more detailed analysis taking into account experimental cuts in the transverse momentum and rapidity of positron as well as the tagging of the forward $Pb^{81^+}$ ion. Currently, we are analyzing how to implement the double particle production via DSM in one of the  Monte Carlo used to simulate  UPHICs, which is the first step to derive more realistic predictions for the processes discussed in this letter and in  Refs. \cite{DSM,Azevedo:2022ozu,Azevedo:2023vsz}.

\begin{acknowledgments}
V.P.G. acknowledges very useful discussions with V. Serbo and R. Schicker about the topic discussed in the  present study during the PHOTON conference in Novosibirsk many years ago. This work was partially supported by CNPq, CAPES, FAPERGS, FAPESC and  INCT-FNA (Process No. 464898/2014-5).

\end{acknowledgments}

\hspace{1.0cm}

\end{document}